\documentclass{article}
\usepackage{fleqn,ams}
\usepackage{amssymb}
\usepackage{verbatim}

\def\stackunder#1#2{\mathrel{\mathop{#2}\limits_{#1}}}
\newcommand{\Per}[1]{#1_{\perp}}
\newcommand{\Par}[1]{#1_{\parallel}}
\newcommand{\epar}{\varepsilon + \Par{p}}

\newcommand{\parminper}{\Par{p} - \Per{p}}
\newcommand{\parminpero}{\Paro{p} - \Pero{p}}
\newcommand{\epso}{\stackrel{0}{\varepsilon}}
\newcommand{\Pero}[1]{\Per{\stackrel{0}{#1}}}
\newcommand{\Paro}[1]{\Par{\stackrel{0}{#1}}}

\newcommand{\po}{\stackrel{0}{p}}
\newcommand{\const}{{\mbox Const}}
\newcommand{\sqtwo}{\sqrt{2}}
\newcommand{\sqtwomin}{\frac{1}{\sqtwo}}

\begin{document}

\begin{center}

{\bf Gravimagnetic shock waves in the anisotropic plasma}\\[12pt]
      Yu.G.Ignatyev,  D. N. Gorokhov\\
    Kazan State Pedagogical
       University,\\Mezhlauk str., 1,
Kazan 420021, Russia

\begin{abstract}The  relativistic magnetohydrodynamic equations for the
anisotropic  magnetoactive plasma are obtained and accurately
integrated in  the  plane gravitational wave metrics. The
dependence of the induction mechanism of the  gravimagnetic  shock
waves  on the degree of the magnetoactive plasma anisotropy is
analyzed.\end{abstract}
\end{center}

\section{Introduction}

In \cite{gmsw} on the basis of an exact solution of the
relativistic  magnetohydrodynamic (RMHD) equations on the
back\-ground  of the  plane  gravitational  wave  (PGW) metrics a
new  class of relativistic essentially non-linear phenomena was
found, which arise in a highly magnetized  plasma under the PGW
influence,  was named as {\it gravimagnetic shock  waves}  (GMSW).
The  essence  of the GMSW phe\-no\-me\-non is that the highly
magnetized plasma
\begin{equation}\label{1}
\alpha^{2}=\frac{\Per{H}^{2}}{4\pi(\varepsilon_{0}+p_{0})}  \gg 1\,,
\end{equation}
(where $\Per{H}$ is the magnetic field intensities component
perpendicular to the PGW propagation direction, $\varepsilon_0$,
$p_0$ are the imperturbed energy density and plasma pressure whithout
the magnetic field) very actively reacts even on a week PGW  by
rather large values of {\it the second  parameter of GMSW}:
\begin{equation}\label{2}
\Upsilon \equiv 2 \beta_0 \alpha^2 > 1\,,
\end{equation}
where $\beta_0$ - is the maximum PGW amplitude.

In \cite{gmsw2} on the basis of the  energy balance model
of the plasma and PGW it was shown, that under the condition
(\ref{2})  the PGW energy is practically fully transformed into
the magnetoactive plasma acceleration (par excellence in the
PGW propagation direction) and into creating a shock  wave
with high densities of the plasma energy and the magnetic
field. In this case the sublight velocity of the plasma motion
in pointed out directionis is achieved.

The essential change of the electromagnetic radiation characteristics
of the plasma in GMSW was used in \cite{gmsw2} in order to creat a new
experimental test for the detection of the gravitational pulsars
radiation. In particular, in \cite{detector} it was shown that, the so-called
giant impulses in the radio radiation of the pulsar NP 0532 may be
explained by the GMSW mechanism.

In the mentioned above papers the locally isotropic plasma was studied
($\Per{p}=\Par{p}=p$); the anisotropy was created by the exclusively
magnetic field. In the strong magnetic fields due to the
magnetobremsstrahlung the local thermodynamic equilibrium (LTE)
of the plasma is destroyed. Therefore generally writing:
\begin{equation}
\label{3}\Per{p}\not=\Par{p}\,.
\end{equation}

The present paper is devoted to the investigation of the degree
influence of the plasma anisotropy on the forming of
the electro-magnetic  reaction of the plasma  to the PGW.
The relativistic magnetohydrodynamic equations are placed into the base
of the theoretical model of the magnetoactive plasma description
in the PGW field \cite{gmsw}.
These equations, as it follows from \cite{probl}, can be also
obtained from the kinetic  equations  to the  collisionless  plasma  in
a drift approximation:
\begin{equation}\label{4}
\omega\ll\omega_{B}\,,
\end{equation}
where $\omega$ is the PGW frequency , $\omega_{B}=e H/m_{e}$
is the Larmor's frequency.  Throughout the paper  the metrics signature
$  (-1,-1,-1,+1) $ is used and the fundamental system  of
units: $G=c=\hbar=1$ are used.

\section{RMHD equations}
\subsection{RMHD equations for an arbitrary \newline
structure of the plasma MET}

Generally the RMHD equations are obtained from the conservation law of the
whole energy-momentum tensor (MET) of the plasma  and electromagnetic field:
\begin{equation}\label{5}
(\stackrel{p}{T^{ij}}+\stackrel{f}{T^{ij}})_{, j}=0
\end{equation}
and from the first group  of the Maxwell equations :
\begin{equation}\label{6}
\stackrel{\ast}{F^{ij}}_{, j}=0
\end{equation}
if the Maxwell tensor satisfies  the following requirements \cite{gmsw}:
\begin{equation}\label{7}
F^{ij} \stackrel{\ast}{F}_{ij}=0 \,;
\end{equation}
\begin{equation}\label{8}
F^{ij} F_{ij}=2 H^2 > 0\,.
\end{equation}
Then it turns out that the timelike eigenvectors of the plasma MET and
of the elactromagnetic field MET coincide, and moreover this vector
of  {\it the plasma dynamic velocity}, $v^i$, is  the eigenvector of
the Maxwell tensor simultaneously: \footnote{In \cite{gmsw} this
condition was called "condition of the magnetic field embedding in the
plasma".}
\begin{equation} \label{9}
F_{ij} v^{j}=0\,.
\end{equation}
Analogical to (\ref{9}) contraction of the velocity vector  of
the dual Maxwell tensor is the magnetic field intensity vector:
\begin{equation}\label{10}
H_i= v^{k}\stackrel{\ast}{F_{ki}}\,,
\end{equation}
This vector is   spacelike:
\begin{equation}\label{11}
(H, H)= - H^2
\end{equation}
and orthogonal to the plasma velocity vector:
\begin{equation} \label{11.a}
(v,H)=0\,.
\end{equation}

Besides, Eq.(\ref{9}) follows from  Eqs.(\ref{5}) - (\ref{8}) (\cite{gmsw})
the second group of the Maxwell equations is:
\begin{equation}\label{12}
F^{ij}_{, j}=- 4\pi J^i_{dr}
\end{equation}
with the spacelike {\it drift current}:
\begin{equation}  \label{13}
J^{i}_{dr}=
- \frac{2 F^{ik} \stackrel{p}{T^{l}}_{k,l}} {F_{jm} F^{jm}}\,;
\end{equation}
and the differential relations:
\begin{equation}\label{14}
v^{i}\stackrel{p}{T^{k}}_{i, k} =0\,,
\end{equation}
\begin{equation} \label{15}
H^{i}\stackrel{p}{T^{k}}_{i, k} =0\,.
\end{equation}

Let us also note useful  differencial indentities following from
Eqs.(\ref{6}) - (\ref{10}) \cite{gmsw}:
\begin{equation} \label{16}
v^{i} H^{k}_{, k} + v^{i}_{, k} H^{k} - v^{k}_{, k} H^{i} -
v^{k} H^{i}_{, k} =0\,;
\end{equation}
\begin{equation} \label{17}
-v_{i, k} H^{i} v^{k} =H_{i, k} v^{i} v^{k}=H^{k}_{, k}\,;
\end{equation}
\begin{equation} \label{18}
H_{i, k} v^{i} H^{k}= -v_{i,k} H^{i} H^{k}=H (H v^{k}) _{, k}\,.
\end{equation}

The pointed above relations represent a complete set of the algebraic and
differential consequences  (\ref{5})  -  (\ref{8}), and at the
the plasma ETM fixating by the set they are a set
of the plasma relativistic  megnetohydrodynamic  equations in
the gravitational field.

\subsection{RMHD equations for the anisotropic plasma}

Let introduce a single spacelike  vector $h^i$ \cite{gmsw}:
\begin{equation} \label{19}
h^i = \frac{H^i}{H}\,;\qquad (h, h)=- 1\,.
\end{equation}
Two independent vectors $v^i$ and $h^i$  and the metrics tensor $ g^{ij}$
define the following algebraic structure of the anisotropic plasma MET:
\begin{equation} \label{20}
\stackrel{p}{T^{ij}} = \left (\varepsilon + \Per{p}\right) v^i v^j -
\Per{p}g^{ij} + \left(\Par{p} -\Per{p} \right) h^i h^j\,,
\end{equation}
and the timelike  vector of velocity satisfies normalization relation:
\begin{equation} \label{20.a}
(v, v)=1\,.
\end{equation}
The MET trace (\ref{20}) is equal to:
\begin{equation} \label{21}
\stackrel{p}{T}=\varepsilon -\Par{p} - 2\Per{p}\,,
\end{equation}
therefore, in the consequence of the known theorem of the virial (see, for
instance \cite{land}) the following condition must take place:
\begin{equation} \label{22}
\Par{p} + 2 \Per{p} \leq \varepsilon\,.
\end{equation}
In this case th electromagnetic field MET is equal to (\cite{gmsw}) :
\begin{equation} \label{22.a}
\stackrel{f}{T}_{ij} = \frac{1}{8\pi} \left (2 H^2 v_i v_j - 2 H_i H_j
- g_{ij} H^2 \right) \,.
\end{equation}

Substituting  the plasma ETM written in the form  (\ref{20}) in
(\ref{14}) and (\ref{15}) and using the indentities
(\ref{16}) - (\ref{18}) we obtained equations:
\begin{equation} \label{23}
\varepsilon_{, i} v^i+(\epar) v^i_{,i}
+ (\parminper) v^i (\ln H)_{,i} = 0\,;
\end{equation}
\begin{equation} \label{24}
(\epar) H^i_{, i} - (\parminper) H^i (\ln H)_{, i}+(\Par{p}) _{, i} H^i
= 0\,.
\end{equation}

\section{ The RMHD equations solution \newline in PGW metrics}
\subsection{ PGW metrics  and the initial conditions}
The PGW metrics has the form \cite{torn}:
\begin{equation} \label{25}
d s^{2} = 2 du dv - L^{2} [e^{2\beta} (dx^{2}) ^{2} +
e^{-2\beta} (dx^{3}) ^{2}]\,,
\end{equation}
where $\beta (u)$ is the amplitude , and $L (u)$ is the PGW
background factor;  $u= \frac{1}{\sqrt{2}} (t - x^{1})$  is the
retarded time, $v = \frac{1}{\sqrt{2}} (t+x^{1})$ is the advanced.
The metrics (\ref{25}) admits  the  group  of  motions
$G_{5}$, associated with three linearly independent  Killing vector:
\begin{equation} \label{25.a}
\stackunder{(1)}{\xi^{i}} = \delta^{i}_{v}\,; \qquad
\stackunder{(2)}{\xi^{i}} = \delta^{i}_{2}\,; \qquad
\stackunder{(3)}{\xi^{i}}=\delta^{i}_{3}\,.
\end{equation}
Let in the PGW absence ($u\leq 0$):
\begin{equation} \label{26}
\beta (u _{\mid u \leq 0} = 0; \qquad L (u)_{\mid u \leq 0}=1,
\end{equation}
the plasma is homogeneous and at rest:
$$ v^{v}_{\mid u \leq 0} = v^{u}_{\mid u \leq 0}= \frac{1}{\sqrt{2}};
\qquad v^{2}=v^{3}=0;
$$
$$
\varepsilon_{\mid u \leq 0}=\varepsilon_{0};
$$
\begin{equation} \label{27}
(\Par{p}) _{\mid u \leq 0} = \Paro{p}; \qquad
(\Per{p}) _{\mid u \leq 0} = \Pero{p}\,,
\end{equation}
 nd the homogeneous magnetic field is directed in the plane
$\lbrace x^{1},x^{2} \rbrace $:
$$
H_{1 \mid u \leq 0}=H_{0} \cos\Omega\,; \qquad
H_{2 \mid u \leq 0}=H_{0} \sin\Omega\,;
$$
\begin{equation}  \label{28}
H_{3 \mid u \leq 0} = 0\,;\qquad E_{i \mid u \leq 0} = 0\,,
\end{equation}
where $\Omega$ is the angle between the axis $0x^{1}$ (the PGW
propogation direction) and  the magnetic field direction ${\bf H}$.
The conditions (\ref{28}) correspond to the vector potential:
$$
A_{v} = A_{u} = A_{2} = 0\,;
$$
\begin{equation} \label{29}
A_{3} = H_{0} (x^{1} \sin\Omega - x^{2} \cos\Omega);
\qquad (u\leq 0).
\end{equation}

In \cite{gmsw} it is shown that in consecuence of Eqs.(\ref{12}) and
(\ref{13}) in the PGW component presence $A_3$ of the vector potential
takes the form:
\begin{equation} \label{30}
A_3 = - H_{0} x^{2} \cos\Omega +
\frac{1}{\sqrt{2}}  H_{0} [v - \psi(u)] \sin\Omega\,,
\end{equation}
where $\psi(u)$ is an arbitrary differented function satisfying
the initial condition:
\begin{equation} \label{31}
\psi_{|u\leq 0}=u\,.
\end{equation}
The components of the magnetic field intensity vector
corresponding to the vector potential (\ref{30}) are equal to:
$$
H_v = - H_0 L^{-2} (v_v \cos\Omega+\frac{1}{\sqrt{2}}v_2\sin\Omega)\,;
$$
$$
H_u = H_0 L^{-2} (v_u \cos\Omega - \frac{1}{\sqrt{2}}\psi'v_2
\sin\Omega)\,;
$$
\begin{equation} \label{32}
H_2 = - \frac{1}{\sqrt{2}} H_0 e^{2\beta} \sin\Omega (v_u+v_v\psi');
\qquad H_3 = 0\,,
\end{equation}
and the embedding conditions (\ref{9}) are reduced to one
equation \cite{gmsw}:
\begin{equation} \label{32.a}
\frac{1}{\sqrt{2}} (v_v \psi' - v_u) \sin\Omega+v^2 \cos\Omega = 0\,.
\end{equation}
In this case the scalar (\ref{11}) is equal to:
\begin{equation} \label{33}
H^2 = H^{2}_{0} \left (\frac{\cos^{2}\Omega}{L^{4}}+
\frac{\sin^{2}\Omega}{L^{2}} \psi' e^{2\beta} \right)\,.
\end{equation}
In \cite{gmsw} it is shown that the relation of the velocity vector
normalization (\ref{20.a}) by means of Eqs.(\ref{32}) and (\ref{33})
may be  represented in the equivalent form:
\begin{equation} \label{34}
(v_v \cos\Omega + \frac{1}{\sqrt{2}} v_2 \sin\Omega)^2 =
\frac{H^2}{H^2_0} v^2_v L^4 - \frac{\sin^2\Omega}{2} L^2 e^{2\beta}\,.
\end{equation}

\subsection{Integrals of the motion}
In the consequence of the existence of the motions (\ref{25.a})
the conservation laws of the whole plasma MET in the PGW field
have the following integrals \cite{gmsw}:
\begin{equation} \label{35}
L^2 \stackunder{(a)}{\xi^i} T_{vi} = C_a =\const\,; \quad
(a = \overline{1, 3})\,,
\end{equation}
of which only the first two are non-trivial. In \cite{gmsw} it was
shown that the consequence of these laws is equivalent to the single
non-trivial Maxwell equation (\ref{12}). Substituting in the integrals
(\ref{35}) the expressions to the MET of the plasma and electromagnetic
field (\ref{20}) and (\ref{22.a}), and using the  relations
(\ref{32}) - (\ref{34}) and also the initial conditions (\ref{26})
and (\ref{27}) we reduce non-trivial integrals to form:
$$
2 L^2 v^2_v (\epar) - e^{2\beta}\frac{(\Pero{H})^2}{H^2} (\parminper) =
$$
\begin{equation} \label{34}
= (\epso + \po) \Delta (u)\,;
\end{equation}
$$
L^2 (\epar) v_v v_2 + \sqtwomin e^{2\beta} \cos\Omega \sin\Omega
\frac{H^2_0}{H^2} (\parminper) =
$$
\begin{equation} \label{35}
= (e^{2\beta} - 1) \frac{H^2_0 \cos\Omega \sin\Omega}{4\sqtwo\pi}+
\sqtwomin \cos\Omega \sin\Omega (\parminpero)\,,
\end{equation}
where:
\begin{equation} \label{35.a}
\po = \cos^2\Omega \Paro{p} + \sin^2\Omega \Pero{p}\,;
\end{equation}
\begin{equation} \label{35.b}
(\Pero{H})^2 = H^2_0 \sin^2\Omega
\end{equation}
and the so-called {\it GMSW governing function} is in\-tro\-du\-ced:
\begin{equation} \label{36}
\Delta (u) = \Bigl[ 1 - \alpha^2 (e^{2\beta} - 1)\Bigr]\,,
\end{equation}
it differs from analogous function to the isotropic plasma \cite{gmsw},
\cite{mistake} by the dimensionaless parameter $\alpha^2$ determination:
\begin{equation} \label{37}
\alpha^2 = \frac{(\Pero{H})^2}{4 \pi (\epso + \po)}\,.
\end{equation}

Solving Eqs.(\ref{34}) and (\ref{35}) relatively $v_v$ and
$v_2$ we get the expressions for the coordinates of the velocity
vector via scalar: $\varepsilon $, $\Par{p}$, $\Per{p}$, $\psi$ and
the explicit functions of retarded time:
\begin{equation} \label{38}
v^2_v = \frac{\epso + \po}{2 L^2 (\epar)} \Delta(u) +
e^{2\beta} \frac{\parminper}{2 L^2 (\epar)} \frac{ (\Pero{H}) ^2}{H^2}\,;
\end{equation}
$$
\frac{v_2}{v_v} = \sqtwomin \sin\Omega \cos\Omega\, \times
$$
\begin{equation} \label{39}\displaystyle
\left\{\frac{(e^{2\beta} - 1) \frac{H^2_0}{4 \pi} - \left[ e^{2\beta}
\frac{H^2_0}{H^2} (\parminper) (\parminpero) \right] }
{\left[ (\epso + \po) \Delta(u) + e^{2\beta} \frac{(\Pero{H})^2}{H^2}
(\parminper) \right]} \right\}\,.
\end{equation}

By means of (\ref{38}) and (\ref{39}) from  the relation of the
velocity vector normalization (\ref{20.a}) we obtain the coordinate
$v_u$ of this vector:
\begin{equation} \label{39.a}
\frac{v_u}{v_v} = \frac{1}{2} \left[ \frac{e^{ - 2\beta}}{L^2}
\left( \frac{v_2}{v_v} \right)^2 + \frac{1}{v^2_v} \right] \,,
\end{equation}
and from the embedding condition (\ref{32. a}) let us find  the value of
the potential derivative $\psi'$:
\begin{equation} \label{39.b}
\psi'= \frac{v_u}{v_v} + \frac{e^{- 2\beta}}{L^2} {\mbox ctg}\Omega
\frac{v_2}{v_v}\,,
\end{equation}
due to it the scalar $H^2$ is deturmined from the relation (\ref{33}).

Using the integrals (\ref{34}) and (\ref{35}) it can be shown that from the
two differencial equations (\ref{23}) and (\ref{24}) only one is independent.
Let Eq.(\ref{23}) be an equation of this kind, which in the PGW metrics
takes the form:
\begin{equation} \label{40}
L^2 \varepsilon' v_v + (\epar) (L^2 v_v)'+
\frac{1}{2}L^2 (\parminper) v_v (\ln H^2)'=0\,.
\end{equation}

Thus, in the final analysis the rest equation (\ref{40}) represents
itself a differential equation for the three unknown scalar functions:
$\varepsilon$,  $\Par{p}$ and $\Per{p}$. Such sub-difiniteness
is the known consequence of the incompleteness of the
hydrodinamic plasma description. For the solution to this
equation it is necessary to impose two supplementary
relations between  the funct1ions  $\varepsilon$,  $\Par{p}$ and
$\Per{p}$, i.e., to introduce state equation of the form:
\begin{equation} \label{41}
\Par{p} = f(\varepsilon)\,; \quad \Per{p}=g(\varepsilon)\,.
\end{equation}

\section{Barothropic state equation}
\subsection{General expressions}
Let us study the barothropic behaviour of the anisotropic plasma, when
the relations (\ref{41}) are linear:
\begin{equation} \label{42}
\Par{p} = \Par{k} \varepsilon\,; \quad \Per{p} = \Per{k} \varepsilon\,,
\end{equation}
moreover due to Eq.(\ref{22}) the constant coefficients $\Par{k}$ and
$\Per{k}$ satisfy the inequality:
\begin{equation} \label{43}
\Par{k} +2 \Per{k} \leq 1\,,
\end{equation}
in the consequence of which it is always:
\begin{equation} \label{43.a}
\Par{p} \leq 1\,; \qquad \Per{p} \leq \frac{1}{2}\,.
\end{equation}
The equation (\ref{40}) at the relations (\ref{42}) is easily
integrated, and we get one more integral:
\begin{equation} \label{44} \displaystyle
\varepsilon (\sqtwo L^2 v_v)^{(1+\Par{k})} H^{(\Par{k} \Per{k})}
= \epso H_0^{(\Par{k} - \Per{k})}\,.
\end{equation}
Thus, formally the problem is solved, as it was reduced to solving
algebraic equations set which, however, is still very complicated for
its solving and analyzing. The solution of the problem is essentially
defined by the three non-dimensional parameters: $\cos\Omega$,
$\Per{k}$ and $\Par{k}$. Below we shall investigate the private values
of these parameters.

\subsection{Transverse propagation of the PGW}
In this case $\cos\Omega=0$, and from (\ref{39}) it followes immediately:
\begin{equation} \label{45}
v_2=0\,.
\end{equation}
Then (\ref{20.a}), (\ref{32.a}) and (\ref{33}) give:
\begin{equation} \label{46} \displaystyle
v_u = \frac{1}{2v_v}\,; \qquad\psi'= \frac{1}{2v^2_v}\,;
\end{equation}
\begin{equation} \label{47} \displaystyle
H = \frac{H_0 e^{\beta}}{\sqtwo L v_v}\,,
\end{equation}
and the subatitution of Eq.(\ref{47}) in Eq.(\ref{38}) leads to the
result:
\begin{equation} \label{48}
v^2_v = \frac{1}{2} \frac{\epso}{L^2 \varepsilon} \Delta (u)\,.
\end{equation}
Substituting (\ref{47}) and (\ref{48}) in (\ref{44}), we obtain
a closed equation for the veriable $\varepsilon$, solving it we finally
get:
\begin{equation} \label{49} \displaystyle
\varepsilon = \epso \left[ \Delta^{1+\Per{k}} L^{2(1+\Par{k})}
e^{2\beta (\Par{k} -\Per{k})} \right]^{ - \Per{g}}\,;
\end{equation}
\begin{equation} \label{50} \displaystyle
v_v = \frac{1}{\sqtwo} \left[ \Delta L^{\Par{k}+\Per{k}}
e^{\beta (\Par{k} -\Per{k})} \right]^{\Per{g}}\,;
\end{equation}
\begin{equation} \label{51} \displaystyle
H = H_0\left[ \Delta L^{(1+\Par{k})}
e^{- \beta (1\Par{k})} \right]^{ - \Per{g}}\,;
\end{equation}
where
\begin{equation} \label{52}
\Per{g} = \frac{1}{1 - \Per{k}} \in [1, 2]\,.
\end{equation}

In particular, for the ultrarelativistic plasma with the zero
longitudinal pressure:
\begin{equation} \label{52.a}
\Par{k} \rightarrow 0\,; \quad \Per{k} \rightarrow \frac{1}{2}
\end{equation}
we get from Eqs.(\ref{49}) - (\ref{52}):
\begin{equation} \label{52.b}
v_v = \sqtwomin L \Delta^2 e^{ -\beta}\,;
\end{equation}
\begin{equation} \label{52.c}
\varepsilon = \epso L^{ - 4} \Delta^{ - 3} e^{2\beta}\,; \quad
H = H_0 L^{ - 2} \Delta^{ - 2} e^{2\beta}\,.
\end{equation}
\subsection{Ultrarelativistic plasma with the \newline outbedding
transverse impulse}
In this case:
\begin{equation}  \label{53a}
\Per{k} = 0\,; \quad \Par{k} = 1\,,
\end{equation}
and we obtain:
\begin{equation}  \label{53b}
v_2 = \sqtwomin L e^{\beta}{\mbox ctg} \Omega [1- \Delta]\,;
\end{equation}
\begin{equation} \label{53c}
v_v = \sqtwomin \Delta L e^{\beta}\,;
\end{equation}
\begin{equation}
\varepsilon = \epso \Delta ^{-1} L^{-4} e^{-2 \beta}\,;
\end{equation}
\begin{equation}  \label{54}
H = H_0 L^{-2} \Delta^{-1}\,.
\end{equation}

\section{The plasma anisotropy infuence  on the GMSW
effectiveness}
In \cite{gmsw} it is pointed out that the singular
state arising in the magnetoactive plasma by fulfiling the
condition (\ref{2}) on the hyperspace
\begin{equation} \label{55}
\Delta(u) = 0\,,
\end{equation}
is destroyed by the back action of the magnetoactive plasma on the PGW,
that leads to an effective absorpting the PGW energy by the plasma
and to the restriction on the PGW amplitude.  In \cite{gmsw2} a simple model
of energoballance allowing to describe this process is bilt.
The gravitational wave  described by  the  metrics  (\ref{25})
corresponds to the "effective tensor of the energy impulse"
has alone non-zero component \cite{torn} with the single non-zero component:
\begin{equation} \label{56}
\stackrel{g}{T}_{uu} = \frac{1}{4\pi} (\beta')^2\,.
\end{equation}
Let $\beta_* (u) $ be the PGW vacuum amplitude and $\beta(u)$
be the PGW amplitude with the energy absorption in the plasma.
Then the energy balance equation in a short-wave  approximation
takes the form:
\begin{equation} \label{57}
(\beta'_*)^2 = (\beta')^2 + 4 \pi
(T_{uu} -\stackrel{0}{T}_{uu})\,,
\end{equation}
where $T_{ik}$ is whole MET of the plasma. At the condition (\ref{1})
in case of the transverse PGW propagation direction, when the
GMSW effect is maximum, the equation (\ref{57}) can be written in the
form (see also \cite{gmsw2}):
\begin{equation} \label{58}
{\dot q}_*^2 = {\dot q}^2 + \xi^2V (q)\,,
\end{equation}
where $q=\beta/\beta_0$ the point means derivation
by the veriable $\sqtwo \omega u$, ($\omega$ is the PGW frequency ),
$V(q)$ is the potential function which in a weak PGW takes the form:
\begin{equation} \label{59}
V (q) = \Bigl[ \Delta(q) \Bigr]^{- 4\Per{g}} - 1\,,
\end{equation}
and $\xi^2$ is the so-called {\it first GMSW parameter} \cite{gmsw2}:
\begin{equation} \label{60} \displaystyle
\xi^2 = \frac{H^2_0}{4 \beta^2_0 \omega^2}\,.
\end{equation}
From (\ref{58}) we obtain the minimally possible value of the governing
function:
\begin{equation} \label{61} \displaystyle
\Delta_{min} = \left( \frac{1}{\xi^2}+1\right)^{ - \Per{\gamma}}\,,
\end{equation}
where:
\begin{equation} \displaystyle
\Per{\gamma} = \frac{1}{4\Per{g}} = \frac{1 -\Per{k}}{4} \leq \frac{1}{4}\,.
\end{equation}
In this case the maximally achievable density of the magnetic field energy
is equal to:
\begin{equation} \label{62} \displaystyle
\left( \frac{H^2}{8\pi} \right)_{max} = \frac{H^2_0}{8\pi}
\sqrt{1+\frac{1}{\xi^2}}
\end{equation}
and generally does not depend on the state equations of
the plasma (\ref{41}). The plasma velocity in the GMSW is
also turns out to be independent on the equation of state.
Maximum density of the plasma energy without the magnetic field turns
out to be dependent on the degree of the plsma anisotropy:
\begin{equation} \label{63} \displaystyle
\varepsilon_{max} = \epso \left( 1+\frac{1}{\xi^2}
\right)^{\frac{1}{4} (1+\Per{k})}
\end{equation}
and is maximum for the ultrarelativistic plasma in the  case  when
lingitudinal pressure is equal to zero.

Thus, the maximum value of the magnitude of the local reaction of the
highly magnetized plasma with the linear equation of state (\ref{41})
does not depend on the plasma anisotropy degree and its
equation of state. Obtained in \cite{gmsw2} the dependence of the maximal
value of local response from the equation of state of plasma, just
like the dependence of pulse duration GMSW from the second its
argument1, are  the  consequences of computed errors arising at
the solution of the equation (\ref{60}) alongside of maximum
of function $q(u)$.  We shall come back to this question in
the following article.

Global magnetobremsstrahlung GMSW-response of the plasma is dependent on
the extents of its anisotropy and equation of state, since
a magneto\-brem\-s\-strah\-lung intensity radiation of one electron is
proportional to the product  of the square of magnetic strength on the
square of lateral momentum $\Per{P}^2$ connected with matching
component of the  pressures, $\Per{p}$. For the detection of this
connection and determi\-ning real equation of state stinstead (\ref{41})
it is necessary the building of kinetic analog GMSW in anisotropic plasma.

\end{document}